\documentclass{pasa}

\def\ASU{$^{1}$}
\def\ANU{$^{2}$}
\def\CSIRO{$^{3}$}
\def\Curtin{$^{4}$}
\def\CfA{$^{5}$}
\def\Haystack{$^{6}$}
\def\MIT{$^{7}$}
\def\NRAO{$^{8}$}
\def\RRI{$^{9}$}

\def\Tata{$^{11}$}
\def\UMelbourne{$^{12}$}
\def\UMichigan{$^{13}$}
\def\USydney{$^{14}$}

\def\UW{$^{16}$}

\def\UWisc{$^{18}$}
\def\Victoria{$^{19}$}
\def\CAASTRO{$^{20}$}
\def\SKASA{$^{21}$}
\def\CALTECH{$^{22}$}
\def\NVIDIA{$^{23}$}
\def\Rhodes{$^{24}$}
\def\Monash{$^{25}$}

\title[The MWA Correlator]{The Murchison Widefield Array Correlator}

\author[Ord et al.]{
S.~M.~Ord\Curtin$^,$\CAASTRO,
B.~Crosse\Curtin,
D.~Emrich\Curtin,
D. Pallot\Curtin,
R.~B.~Wayth\Curtin$^,$\CAASTRO,
M.~A.~Clark\NVIDIA$^,$\CfA$^,$\CALTECH,
S.~E.~Tremblay\Curtin$^,$\CAASTRO,
W.~Arcus\Curtin, 
D.~Barnes\Monash, 
M.~Bell\USydney,
G.~Bernardi\SKASA$^,$\Rhodes$^,$\CfA, 
N.~D.~R.~Bhat,\Curtin,
J.~D.~Bowman\ASU, 
F.~Briggs\ANU,
J.~D.~Bunton\CSIRO, 
R.~J.~Cappallo\Haystack, 
B.~E.~Corey\Haystack, 
A.~A.~Deshpande\RRI, 
L.~deSouza\CSIRO$^,$\USydney,
A.~Ewell-Wice\MIT,
L.~Feng\MIT,
R.~Goeke\MIT,
L.~J.~Greenhill\CfA,
B.~J.~Hazelton\UW, 
D.~Herne\Curtin, 
J.~N.~Hewitt\MIT, 
L.~Hindson\Victoria,
H.~Hurley-Walker\Curtin,
D.~Jacobs\ASU,
M.~Johnston-Hollitt\Victoria,
D.~L.~Kaplan\UWisc, 
J.~C.~Kasper\UMichigan$^,$\CfA, 
B.~B.~Kincaid\Haystack, 
R.~Koenig\CSIRO, 
E.~Kratzenberg\Haystack, 
N.~Kudryavtseva\Curtin,
E.~Lenc,\USydney$^,$\CAASTRO
C.~J.~Lonsdale\Haystack, 
M.~J.~Lynch\Curtin, 
B.~McKinley\ANU,
S.~R.~McWhirter\Haystack,
D.~A.~Mitchell\CSIRO$^,$\CAASTRO, 
M.~F.~Morales\UW, 
E.~Morgan\MIT, 
D.~Oberoi\Tata, 
A.~Offringa,\ANU$^,$\CAASTRO,
J.~Pathikulangara\CSIRO, 
B.~Pindor\UMelbourne,
T.~Prabu\RRI, 
P.~Procopio\UMelbourne,
R.~A.~Remillard\MIT, 
J.~Riding\UMelbourne,
A.~E.~E.~Rogers\Haystack, 
A.~Roshi\NRAO, 
J.~E.~Salah\Haystack, 
R.~J.~Sault\UMelbourne, 
N.~Udaya~Shankar\RRI, 
K.~S.~Srivani\RRI, 
J.~Stevens\CSIRO,
R.~Subrahmanyan\RRI$^,$\CAASTRO, 
S.~J.~Tingay\Curtin$^,$\CAASTRO, 
M.~Waterson\Curtin$^,$\ANU,
R.~L.~Webster\UMelbourne$^,$\CAASTRO, 
A.~R.~Whitney\Haystack, 
A.~Williams\Curtin, 
C.~L.~Williams\MIT, 
J.~S.~B.~Wyithe\UMelbourne$^,$\CAASTRO
\\
\\
$^{1}$Arizona State University, Tempe, USA\\
$^{2}$The Australian National University, Canberra, Australia\\
$^{3}$CSIRO Astronomy and Space Science, Australia\\
$^{4}$International Centre for Radio Astronomy Research (ICRAR), Curtin University, Perth, Australia\\
$^{5}$Harvard-Smithsonian Center for Astrophysics, Cambridge, USA\\
$^{6}$MIT Haystack Observatory, Westford, MA, USA\\
$^{7}$MIT Kavli Institute for Astrophysics and Space Research, Cambridge, USA\\
$^{8}$National Radio Astronomy Observatory, Charlottesville, USA\\
$^{9}$Raman Research Institute, Bangalore, India\\
$^{10}$Swinburne University of Technology, Melbourne, Australia\\
$^{11}$National Center for Radio Astrophysics, Pune, India\\
$^{12}$The University of Melbourne, Melbourne, Australia\\
$^{13}$University of Michigan, Ann Arbor, USA\\
$^{14}$University of Sydney, Sydney, Australia\\
$^{15}$University of Tasmania, Hobart, Australia\\
$^{16}$University of Washington, Seattle, USA\\
$^{17}$University of Western Australia, Perth, Australia\\
$^{18}$University of Wisconsin--Milwaukee, Milwaukee, USA\\
$^{19}$Victoria University of Wellington, New Zealand\\
$^{20}$ARC Centre of Excellence for All-sky Astrophysics (CAASTRO)\\
$^{21}$Square Kilometre Array South Africa (SKA SA), Cape Town, South Africa\\
$^{22}$California Institute of Technology, California, USA\\
$^{23}$NVIDIA, Santa Clara, California, USA\\
$^{24}$Department of Physics and Electronics, Rhodes University, Grahamstown, South Africa\\
$^{25}$Monash University, Melbourne, Australia\\
}

\jid{PASA}
\doi{10.1017/pas.\the\year.xxx}
\jyear{\the\year}
\usepackage{graphicx}
\usepackage{booktabs}
\usepackage{animate}
\usepackage{pgfplotstable}

\usepackage[authoryear]{natbib}
\bibpunct{(}{)}{;}{a}{}{,}
\usepackage{dsfont}

\begin{document}
\begin{abstract}

The Murchison Widefield Array (MWA) is a Square Kilometre Array (SKA) Precursor. The telescope is located at the Murchison Radio--astronomy Observatory (MRO) in Western Australia (WA). The MWA consists of 4096 dipoles arranged into 128 dual polarisation aperture arrays forming a connected element interferometer that cross-correlates signals from all 256 inputs. A hybrid approach to the correlation task is employed, with some processing stages being performed by bespoke hardware, based on Field Programmable Gate Arrays (FPGAs), and others by Graphics Processing Units (GPUs) housed in general purpose rack mounted servers. The correlation capability required is approximately 8 TFLOPS (Tera FLoating point Operations Per Second). The MWA has commenced operations and the correlator is  generating 8.3 TB/day of correlation products, that are 
subsequently transferred 700 km from the MRO to Perth (WA) in real-time for storage and offline processing. In this paper we outline the correlator design, signal path, and 
processing elements and present the data format for the internal and external interfaces.
  
\end{abstract}
\begin{keywords}
instrumentation: interferometers, techniques: interferometric
\end{keywords}
\maketitle%
\section{Introduction}

The MWA is a 128 element dual polarisation interferometer, each element is a 4x4 array of analog beam formed dipole antennas. The antennas of each array are arranged in a regular grid approximately 1m apart, and these small aperture arrays are known as tiles.  The science goals that have driven the MWA design and development process are discussed in the instrument description papers\citep{tingay:2013,lonsdale:2009}, and the MWA science paper \citep{bowman:2013}. These are (1) the detection of redshifted 21cm neutral hydrogen from the Epoch of Re-ionization (EoR);  (2) Galactic and extra-Galactic surveys; (3) time-domain astrophysics; (4) solar, heliospheric and ionospheric science and space weather. 

\subsection{Specific MWA Correlator Requirements}

The requirements and science goals have driven the MWA into a compact configuration of 128 dual-polarisation tiles. 50 tiles are concentrated in the 100-m diameter core, with 62 tiles distributed within 750-m  and the remaining 16 distributed up to 1.5-km from the core.  

The combination of the low operating frequency of the MWA and its compact configuration allow the correlator to be greatly reduced in complexity, however this trade-off does drive the correlator specifications. Traditional correlators are required to compensate for the changing geometry of the array with respect to the source in order to permit coherent integration of the correlated products. In the case of the MWA, no such corrections are performed. This drives the temporal resolution specifications of the correlator and forces the products to be rapidly generated in order to maintain coherence. \cite{tingay:2013} list the system parameters and these include a temporal resolution of 0.5~s. 

The temporal decoherence introduced by integrating for 0.5~s without correcting for the changing array geometry; and the requirement to image a large fraction of the primary beam, drive the system channel resolution specification to 10~kHz. It should be noted that this does not drive correlator complexity as it is total processed bandwidth that is the performance driver, not the number of channels. Albeit the number of output channels does drive the data storage and archiving specifications.  The MWA correlator is required to process the full bandwidth as presented by the current MWA digital receivers, which is 30.72~MHz per polarisation.

In summary, in order to meet the MWA science requirements, the correlator is required to correlate 128 dual polarisation inputs, each input is 3072x10kHz in bandwidth. The correlator must present products, integrated for no more than 0.5 seconds at this native channel resolution, to the data archive for storage.

\subsection{Benefits of Software Correlator Implementations}

The correlation task has previously been addressed by  Application Specific Integrated Circuits (ASICs) and Field Programmable Gate Arrays (FPGAs). 
However the current generation of low frequency arrays including the MWA \citep{wayth:2009}, LOFAR \citep{LOFAR:2013}, PAPER  \citep{PAPER:2010} and LEDA \citep{LEDA:2014} have chosen to utilise, to varying degrees, general purpose computing assets to perform this operation. The MWA leverages two technologies to perform the correlation task, a Fourier transformation performed by a purpose built FPGA-based solution, and a cross-multiply and accumulate task (XMAC) utilising the xGPU library\footnote{ xGPU available at: https://github.com/GPU-correlators/xGPU} described by \cite{clark:2011}, also used by the PAPER telescope and the LEDA project. The MWA system was deployed over 2012/13 as described by \cite{tingay:2013}, and is now operational. 

The MWA is unique amongst the recent low frequency imaging dipole arrays, in that it was not designed to utilise a software correlator at the outset. The flexibility provided by the application of general purpose, GPU based, software solution has allowed a correlator to be rapidly developed and fielded. This was required in-order to respond to a changed funding environment, which resulted in a significant design evolution from that initially proposed by \cite{lonsdale:2009} to that described by \cite{tingay:2013}.  Expeditious correlation development and deployment was possible because General Purpose GPU (GPGPU) computing provides a compute capability, comparable to that available from the largest FPGAs, in a form that is much more accessible to software developers. It is true that GPU solutions are more power intensive than FPGA or ASIC solutions, but their compute capability is already packaged in a form factor that is readily available and the development cycle is identical to other software projects. Furthermore the GPU processor lifecycle is very fast and it is also generally very simple to benefit from architecture improvements. For example, the GPU kernel used as the cross-multiply engine in the MWA correlator will run on any GPU released after 2010. We could directly swap out the GPU in the current MWA cluster, replace them with cards from the Kepler series (K20X), and realise a factor of 2.5 increase in performance (and a threefold improvement in power efficiency).

The organisation of this paper begins with a short introduction to the correlation problem, followed by an outline of the MWA correlator design and then a description of the sub-elements of the correlator following the signal path; we finally discuss the relevance of this correlator design to the Square Kilometre Array and there are several Appendices describing the various internal and external interfaces.




\section{The Correlation Problem}

A traditional telescope has a filled aperture, where a surface or a lens is used to focus incoming radiation to a focal point. In contrast, in an interferometric array like the MWA the purpose of a correlator is to measure the level of signal correlation between all antenna pairs at different frequencies across the observing band. These products can then be added coherently, and phased, or focussed, to obtain a measurement of the sky brightness distribution in any direction 

The result  of the correlation operation is commonly called a {\em visibility} and is a representation of the measured signal power (brightness) from the sky on angular scales commensurate with the distance between the constituent pair of antennas. Visibilities generated between antennas that are relatively far apart measure power on smaller angular scales and vice versa. When all visibilities are calculated from all pairs of antennas in the array many spatial scales are sampled (see \cite{TMS} for an extensive review of interferometry and synthesis imaging). When formed as a function of observing frequency ($\nu$), this visibility set forms the {\em cross-power spectrum}. For any two voltage time-series from any two antennas $V_{1}$ and  $V_{2}$ this product can be formed in two ways. First the cross correlation as a function of lag, $\tau$, can be found, typically by using a delay line and multipliers to form the lag correlation between the time-series.
\begin{equation}
(V_{1} \star V_{2})(\tau) = \int_{-\infty}^{\infty} V_{1}(t)V_{2}(t-\tau) dt.
\label{eqn:xcorr}
\end{equation}
The cross power spectrum, $S(\nu)$, is then obtained by application of a Fourier transform:
\begin{equation}
S_{12}(\nu) = \int_{-\infty}^{\infty} (V_{1} \star V_{2})(\tau)e^{-2\pi i \nu\tau} d\tau.
\label{eqn:XF}
\end{equation}
When the tasks required to form the cross power spectrum are performed in this order (lag cross-correlation, followed by Fourier transform) the combined operation is considered an {\em XF} correlator. However the cross correlation analogue of the convolution theorem allows Equation \ref{eqn:XF} to be written as the product of the Fourier transform of the voltage time series from each antenna:
\begin{equation}
S_{12}(\nu) = \int_{-\infty}^{\infty} V_{1}(t) e^{-2\pi i \nu t} dt  \times \int_{-\infty}^{\infty} V_{2}(t) e^{-2\pi i \nu t} dt.
\label{eqn:FX}
\end{equation}
Implemented as described by Equation \ref{eqn:FX} the operation is an {\em FX} correlator.  For large-N telescopes the FX correlator has a large computational advantage. In an XF correlator for an array of $N$ inputs the cross correlation for all baselines requires $\mathcal{O}(N^{2})$ operations for every lag, and there is a one to one correspondence between lags and output channels, F, resulting in $\mathcal{O}(FN^{2})$ operations to generate the full set of lags. The Fourier transform requires a further $\mathcal{O}(F\log_{2}F)$ operations, but this can be performed after averaging the lag spectrum and is therefore inconsequential. For the FX correlator we require $\mathcal{O}(NF\log_{2}F)$ operations  for the Fourier transform of all input data streams, but only $\mathcal{O}(N^{2})$ operations per sample for the cross multiply (although we have F channels the sample rate is now lower by the same factor). Therefore as long as $N$ is greater than $\log_{2}F$ there is a  computational advantage in implementing an FX correlator.  XF correlators have been historically favoured by the astronomy community, at least in real-time applications, as until very recently  $N$ has been small, and there are disadvantages to the FX implementation. The predominant disadvantage is data growth: the precision of the output from the Fourier transform is generally larger than the input, resulting in a data rate increase. There is also the complexity of implementing the Fourier transform in real-time.


\section{The MWA Correlator System}

\begin{figure}
   \includegraphics[width=\columnwidth]{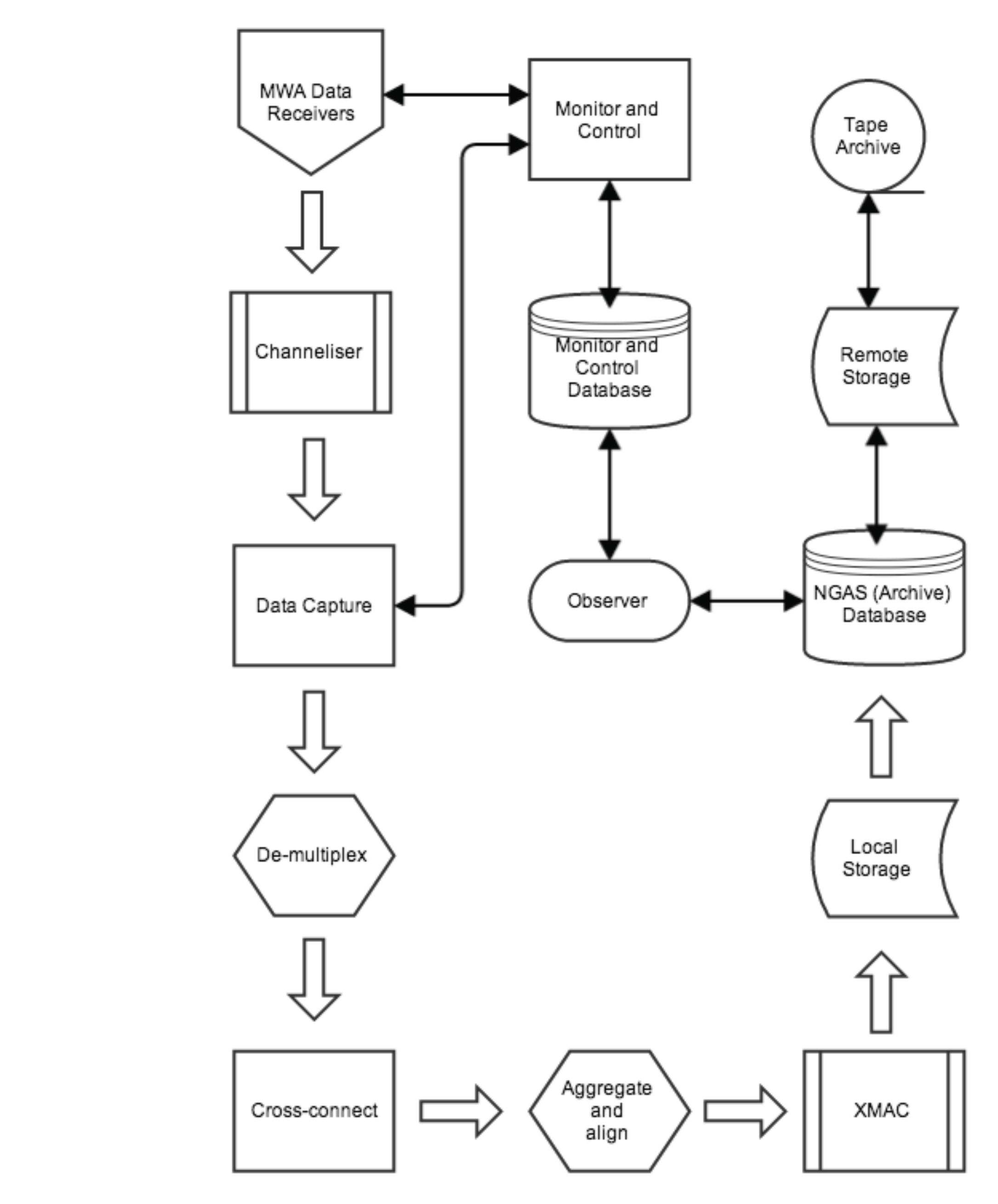} 
   \caption{The MWA signal processing path, following the MWA data receivers, in the form of a flow diagram. }
   \label{flow}
\end{figure}

\begin{figure*}
\includegraphics[width=\textwidth]{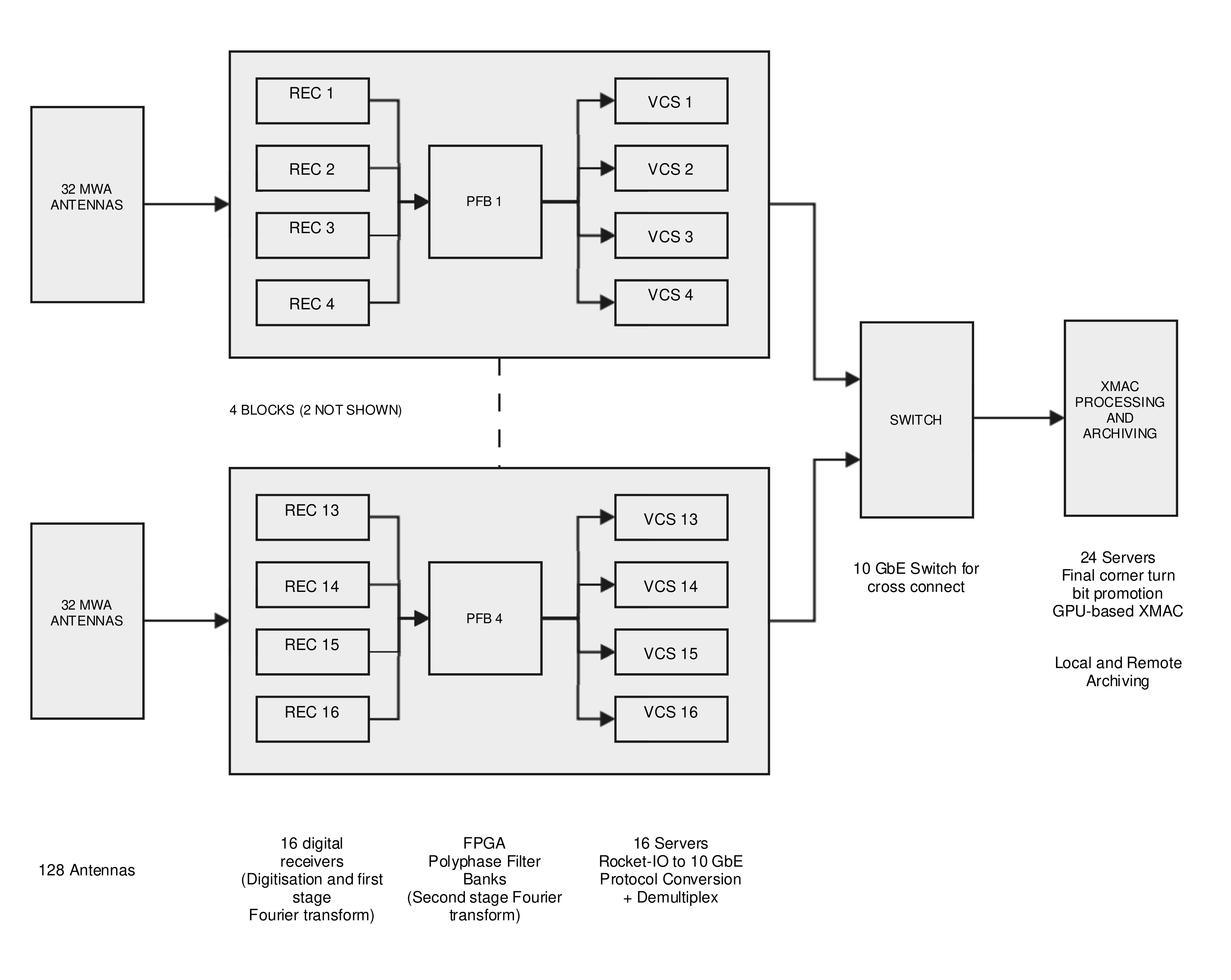}
\caption{The full physical layout of the MWA digital signal processing path. The initial digitalisation and coarse channelisation is performed in the field. The fine channelisation, data distribution and correlation is performed in the computer facility at the Murchison Radio Observatory, the data products are finally archived at the Pawsey Centre in Perth.}
\label{layout}
\end{figure*}

\begin{figure*}
\includegraphics[width=\textwidth]{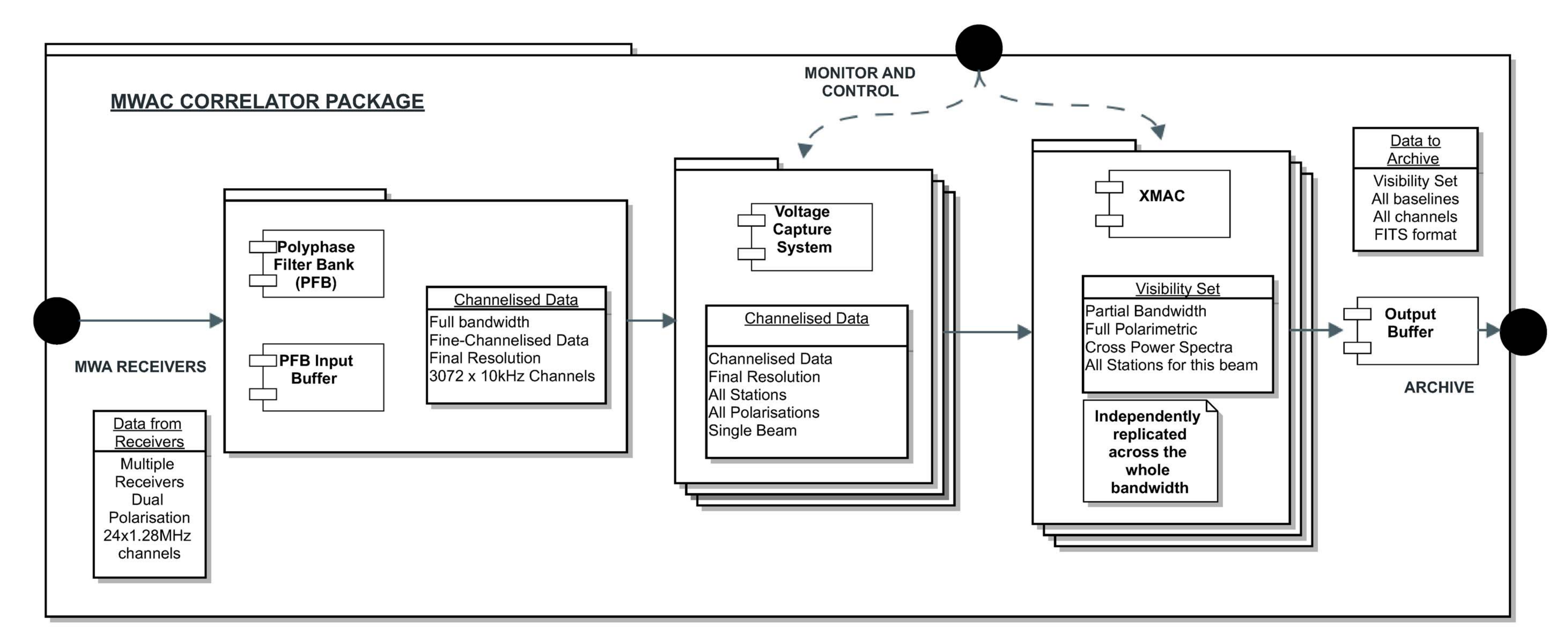}
\caption{The decomposition of the MWA Correlator system demonstrating the relationship between the major sub-elements.}
\label{decomp}
\end{figure*}

The tasks detailed in this section cover the full signal path after digitisation, including fine channelisation, data distribution, correlation and output. In the MWA correlator the F stage is performed by a dedicated channeliser, subsets of frequency channels from all antennas are then distributed to a cluster of processing nodes via an ethernet switch. The correlation products are then assembled and distributed to an archiving system. An outline of the system as a whole is shown in Figures \ref{flow} and \ref{layout}. As shown in Figure \ref{decomp} the correlator is conceptually composed of 4 sub-packages, the Polyphase Filterbank (PFB), that performs the fine channelisation, the Voltage Capture System (VCS), responsible for converting the data transport protocol into ethernet and distributing the data, the correlator itself (XMAC), and the output buffering that provides the final interface to the archive. Figure \ref{layout} presents the physical layout of the system.

\subsection {The System Level View}

The computer system consists of 40 servers, grouped into different tasks and connected by a 10GbE  data network and several 1GbE monitoring, command and control networks. The correlator computer system has been designed for ease of maintenance and reliability. All of the servers are configured as {\em thin-clients} and have no physical hard-disk storage that is utilised for any critical functions - any hard-disk is used for non-critical storage and could be removed (or fail) without limiting correlator operation. All of the servers receive their Operating System through a process known as the PXE (Pre-eXecution Environment) boot process, and import all of their software over NFS (Network File System) and mount it locally in memory. The servers are grouped by task into the VCS (voltage capture system) machines that house the FPGA capture card and the servers that actually perform the cross-multiply-accumulate (XMAC) and output the visibility sets for archiving. Any machine can be replaced without any more effort than is required to physically connect the server box and update an entry in the IP address mapping tables.

\subsection{The MWA Data Receivers: Input to the Correlator}

As described in \cite{receiver} there are 16 digital receivers deployed across the MWA site. Power and fibre is reticulated to them as described by \cite{tingay:2013}. The individual tiles are connected via coaxial cable to the receivers, the receivers are connected via fibre runs of varying lengths to the central processing building of the MRO, where the rest of the signal processing is located.

\subsubsection{The Polyphase Filterbank}

Polyphase Filter Banks (PFBs) are commonly used in digital signal communications as a realisation of a Discrete Fourier Transform (DFT; see  \cite{crochiere:1983} for a detailed explanation). It specifically refers to the formation of a DFT via a method of {\em decimating} input samples into a polyphase filter structure and forming the DFT via the application of a Fast Fourier Transform to the output of the polyphase filter \citep{bunton:2003}. 

As described by \cite{receiver} the first or {\it coarse} PFB is as an 8 tap, 512 point, critically sampled, polyphase filter bank. This is implemented as; an input filtering stage realised by 8x512 point Kaiser windowing function, the output of which is subsampled into a 512 point FFT. This process transforms the 327.68\,MHz input bandwidth into 256, 1.28\, MHz wide, subbands. 

The second stage of the F is performed by four dedicated FPGA based polyphase filter bank boards housed in two ATCA (Advanced Telecommunications Computing Architecture) racks, which implement a 12 tap, 128 point filter response weighted DFT. Because the first stage is critically sampled the fine channels that are formed at the boundary of the coarse channels are corrupted by aliasing.  These channels are processed by the pipeline, but removed during post-processing, the fraction of the band excised due to aliasing is approximately 12\%. The second stage is also critically sampled so there is a similar degree of aliasing present in each 10kHz wide channel. The aliased signal correlates, and the phase of the correlation does not change significantly across the narrow channel, any further loss in sensitivity is negligible.

Development of the PFB boards was initially funded through the University of Sydney and the firmware initially developed by CSIRO. The boards were designed to form part of the original MWA correlator which would share technology with the Square Kilometre Array Molongolo Prototype (SKAMP) \citep{desouza:2007}. Subsequent modifications to the firmware were undertaken at MIT-Haystack and the final boards, with firmware specific to the MWA, were first deployed as part of the 32-tile MWA prototype. 

\subsubsection{Input Format, Skew and Packet Alignment}

The PFB board  input data format is the Xilinx serial protocol RocketIO, although there have been some customisations at a low level, made within the RocketIO CUSTOM scheme. The data can be read by any multi-gigabit-transceiver (MGT) that can use this protocol, which in practice is restricted to the Xilinx FPGA (Vertex 5 and newer). The packet format as presented to the PFB board for the second stage channeliser is detailed in \cite{receiver}. 


A single PFB board processes 12 fibres that have come from 4 different receivers.  The receivers are distributed over the 3~km diameter site and there exists the possibility of there being considerable difference in packet arrival time between inputs that are connected to different receivers.  The PFB boards have an input buffer that is used to align all receiver inputs on packet number 0 (which is the 1 second tick marker see Appendix \ref{VCS}). The buffer is of limited length ($\pm 8$ time-samples, corresponding to several hundred metres of fibre) and if the relative delay between input lines from nearby and distant receivers is greater than this buffer length, the time delay between receivers will be undetermined. We have consolidated the receivers into 4 groups with comparable distances to the central processing building and constrained these groups to have fibre runs of 1380, 925, 515,  and 270 metres.  Each of the 4 groups is allocated to a single PFB (see Figure \ref{layout}). This has resulted in more fibre being deployed than was strictly necessary, but has guaranteed that all inputs to a PFB board will arrive within a packet-time. The limited buffer space available can therefore be utilised to deal with variable delays induced by environmental factors and will guarantee that all PFB inputs will be aligned when presented to the correlator. 



This packet alignment aligns all inputs onto the same 1 second mark but does not compensate for the {\em outward} clock signal delay, which traverses the same cable length, and produces a large cable delay between the receiver groups, commensurate with the differing fibre lengths. The largest cable length difference is 1110 metres, being the difference in length between the shortest and longest fibres and is removed during calibration.

\subsubsection{PFB Output Format}

As the MWA PFB boards were originally purposed as the hardware F-stage of an FPGA based FX correlator, the output of the PFB was never intended to be processed independently. The output format is also the Xilinx serial protocol RocketIO, and the PFB presents eight output data lanes on two CX4 connectors. These connectors generally house four-lane XAUI  as used by Inifiniband and 10Gb Ethernet, but these are simply used as a physical layer. In actuality all of the pins on the CX4 connectors are driven as transmitters, as opposed to receive and transmit pairs, therefore only half of the available pins in the CX4 connectors were being used. We were subsequently able to change the PFB firmware to force all output data onto connector pins that are traditionally transmit pins, and leave what would normally be considered the receive pins unused. We then custom designed breakout cables that split the CX4 into 4 single lane SFP connectors (see Appendix \ref{VCS}).

 
\subsection{The Voltage Capture System}

In order to capture data generated by the PFBs with off-the-shelf hardware, the custom built SFP connectors were then plugged into an off-the-shelf Xilinx based RocketIO capture cards that are housed within linux servers. Each card can capture 2 lanes from the PFB cards, therefore 4 cards were required per PFB, and 16 cards required in total.  A set of 16 machines are dedicated to this voltage capture system (VCS), these are 16 CISCO UCS C240 M3 servers. They each house two Intel Xeon E5-2650 processors, 128 GB of RAM, and 2x1TB RAID5 disk arrays. These machines also mount the Xilinx-based FPGA board, supplied by Engineering Design Team Incorporated (EDT) that is used to capture the output from the PFB. All of the initial buffering and packet synchronisation is performed within these machines. 
\begin{figure*}
\includegraphics[width=\textwidth]{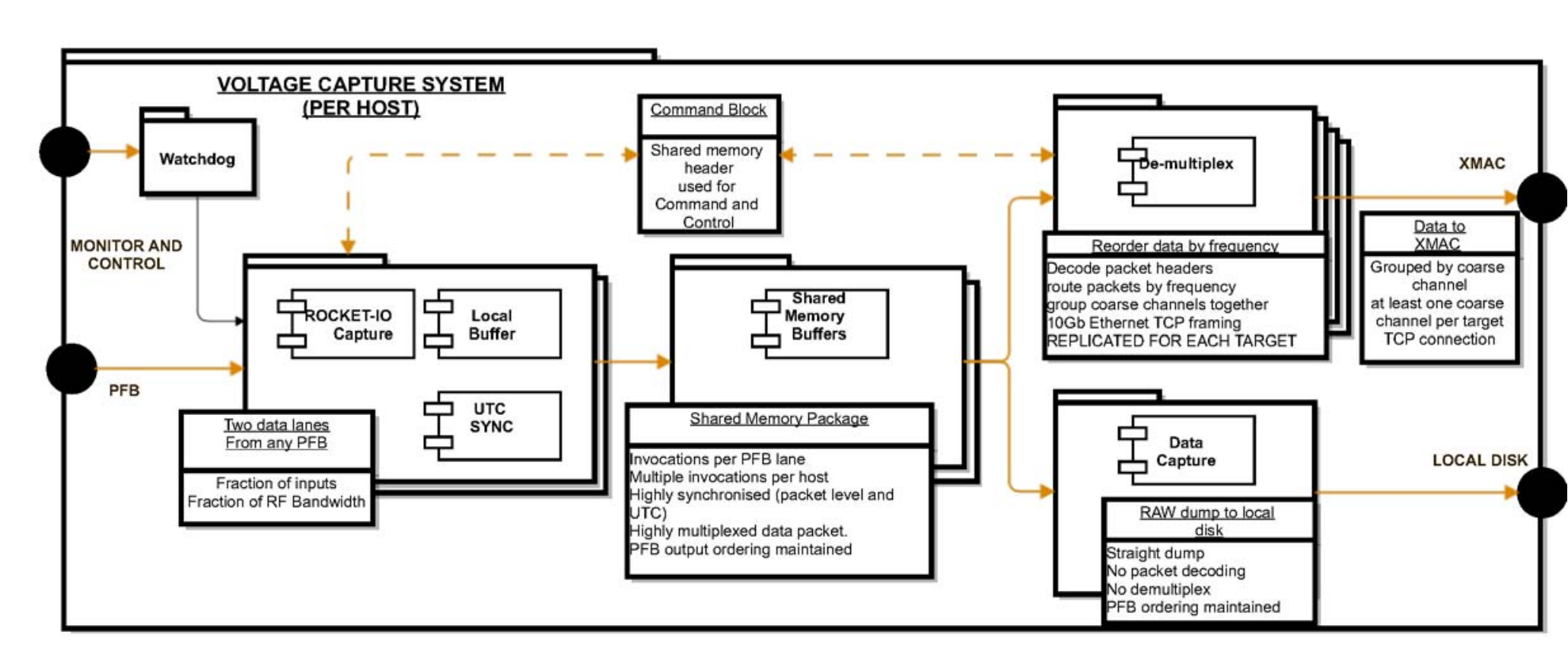}
\caption{The decomposition of the Voltage Capture System. The diagram shows the individual elements that comprise the VCS on a single host. There are 16 VCS hosts operating independently (but synchronously) within the correlator system as a whole.}
\label{capture}
\end{figure*}

\subsubsection{Raw Packet Capture}

The data capture and distribution is enabled by a succession of software packages outlined in Figure \ref{capture}. First, the EDT supplied capture card transfers a raw packet stream from device memory to CPU host memory. The transfer is mediated by the device driver and immediately copied to a larger shared memory buffer. At this stage the data are checked for integrity and aligned on packet boundaries to ensure efficient routing without extensive re-examination of the packet contents. Each of the 16 MWA receivers in the field obtain time synchronisation via a distributed 1 pulse per second (PPS) that is placed into the data stream in the  ``second tick'' field as described in Appendix \ref{VCS}. We maintain synchronisation with this ``tick'' and ensure all subsequent processing within the correlator labels the UTC second correctly. As each machine operates independently the success or failure of this synchronisation method can only be detected when packets from all the servers arrive for correlation, at which point the system is automatically resynchronised if an error is detected. No unsynchronised data are correlated. 

\subsubsection{The Demultiplex}

Data distribution in a connected element interferometer is governed by two considerations; all the inputs for the same frequency channels for each time step must be presented to the correlator; and the correlator must be able to keep up with the data rate. 

As there are four PFB boards (see Figure \ref{layout}), each processing one quarter of the array, each output packet contains the antenna inputs to that PFB, for a subset of the channels for a single time step. Individual data streams from each antenna in time order need to be {\em cross-connected} to satisfy the requirement that a single time step for {\em all antennas} is presented simultaneously for cross correlation.  To achieve this we utilise the fact that  each PFB packet is uniquely identified by the contents of its header (see Appendix \ref{VCS}). We route each packet (actually each block of 2000 packets) to an endpoint based upon the contents of this header, each VCS server sends 128 contiguous, 10kHz channels to the same physical endpoint, from all of the antennas in its allocation. This results in the same 128 frequency channels, from all antennas in the array, arriving at the same physical endpoint. This is repeated for 24 different endpoints -- each receiving a different 128 channels. These endpoints are the next group of servers, the cross multiply and accumulate (XMAC) machines.

This cross-connect is facilitated by a 10\,Gb ethernet switch. In actuality this permits the channels to be aggregated on any number of XMAC servers. We use 24 as it suits the frequency topology of the MWA. Regardless of topology some level of distribution is required to reduce the load on an individual XMAC server in a flexible manner, as compute limitations govern how many frequency channels can be simultaneously processed in the XMAC stage. \cite{clark:2011} demonstrates that the XMAC engine performance is limited to processing approximately ~4096 channels per NVIDIA Fermi GPU  at the 32 antenna level and ~1024 channels at the 128 antenna level. This is well within the MWA computational limitations as we are required to process  3072 channels, and have up to 48 GPU available. However, the benefit of packet switching correlators in general is clear: if performance were an issue we could simply add more GPU onto the switch to reduce the load per GPU without adding complexity. Conversely, as GPU performance improves we can reduce the number of servers and still perform the same task.

Each of the VCS servers is required to maintain 48 concurrent TCP/IP connections, two to each endpoint, because each VCS server captures two lanes from a PFB board . This amounts to 48 open sockets that are being written to sequentially. As there are many more connections than available CPU cores individual threads, tasked to manage each connection, the operation is subject to the linux thread scheduler, which attempts to distribute CPU resources in a {\em round robin} manner. Each thread on each transmit line utilises a small ring buffer to allow continuous operation despite the inevitable device contention on the single 10\,Gb ethernet line. Any such contention causes the threads without access to the interface to block. This time spent in this {\em wait condition} allows the scheduler to redistribute resources. We also employ some context-based thread waiting on the receiving side of the sockets to help the scheduler in its decision making and to even the load across the data capture threads.  We have found that this scheme results in a thread scheduling pattern that is remarkably fair and equitable in the allocation of CPU resources and provided that a reasonably-sized ring buffer is maintained for each data line, all the data are transmitted without loss.

\subsubsection{VCS Data Output}

To summarise, we are switching packets as they are received and they are routed based upon the contents of their header. Therefore any VCS server can process any PFB lane output. Each server holds a single PCI-based capture card, and captures two PFB data lanes. The data from each lane is grouped in time contiguous blocks of 2000 packets (50 milliseconds worth of samples for 16 adjacent 10~kHz frequency channels), from all 32 of the antennas connected to that PFB. The packet header provides sufficient information to uniquely identify the  PFB, channel group, and time block it is, and the packet block is routed on that basis. Further information is given in Appendix \ref{XMAC_Ap}. A static routing table is used to ensure that each XMAC server receives a contiguous block of frequency channels, the precise number of which  is flexible and determined by the routing table, but it is not alterable at runtime. {\footnote {In order to achieve optimal loop unrolling and compile-time evaluation of conditional statements, xGPU requires that the number of channels, number of stations, and minimum integration time are compile-time parameters. }}

\subsubsection{Monitoring, Command, Control and the Watchdog}

Monitor and control functionality within the correlator is mediated by a watchdog process that runs independently on each server. The watchdog launches each process, monitors activity, and checks error conditions. It can restart the system if synchronisation is lost and mediates all start/stop/idle functionality as dictated by the observing schedule. All interprocess command and control, for example propagation of HALT instructions to child processes and time synchronisation information, is maintained in a small block of shared memory that holds a set of key-value pairs in plain text that can be interrogated or set by third-party tools to control the data capture and check status.

 \subsubsection{Voltage Recording}
 
The VCS system also has the capability to record the complete output of the PFB to local disk. The properties and capabilities of the VCS system will be detailed in a companion paper \citep{tremblay:2014}.

\subsection{The Cross-Multiply and Accumulate (XMAC)}

The GPU application for performing the cross-multiply and accumulate is running on 24 NVIDIA M2070 Fermi GPU housed in 24 IBM iDataplex servers. These machines are housed in racks adjacent to the VCS servers and connected via a 10Gb ethernet network. The package diagram detailing this  aspect of the correlator is presented in Figure \ref{xmac} and the packet interface and data format is described in Appendix \ref{GPU_Ap}.
The cross-multiply servers receive data from the 32 open sockets and internally align and unpack the data into a form suitable for the GPU. The GPU kernel is launched to process the data allocation and internally integrates over a user defined length of time. There are other operations possible, such as incoherent beam forming and the raw  dump of data products (or input voltages). These modes will be expanded upon during the development of other processing pipelines and will be detailed in companion papers when the modes are available to the community.

\begin{figure*}
\includegraphics[width=\textwidth]{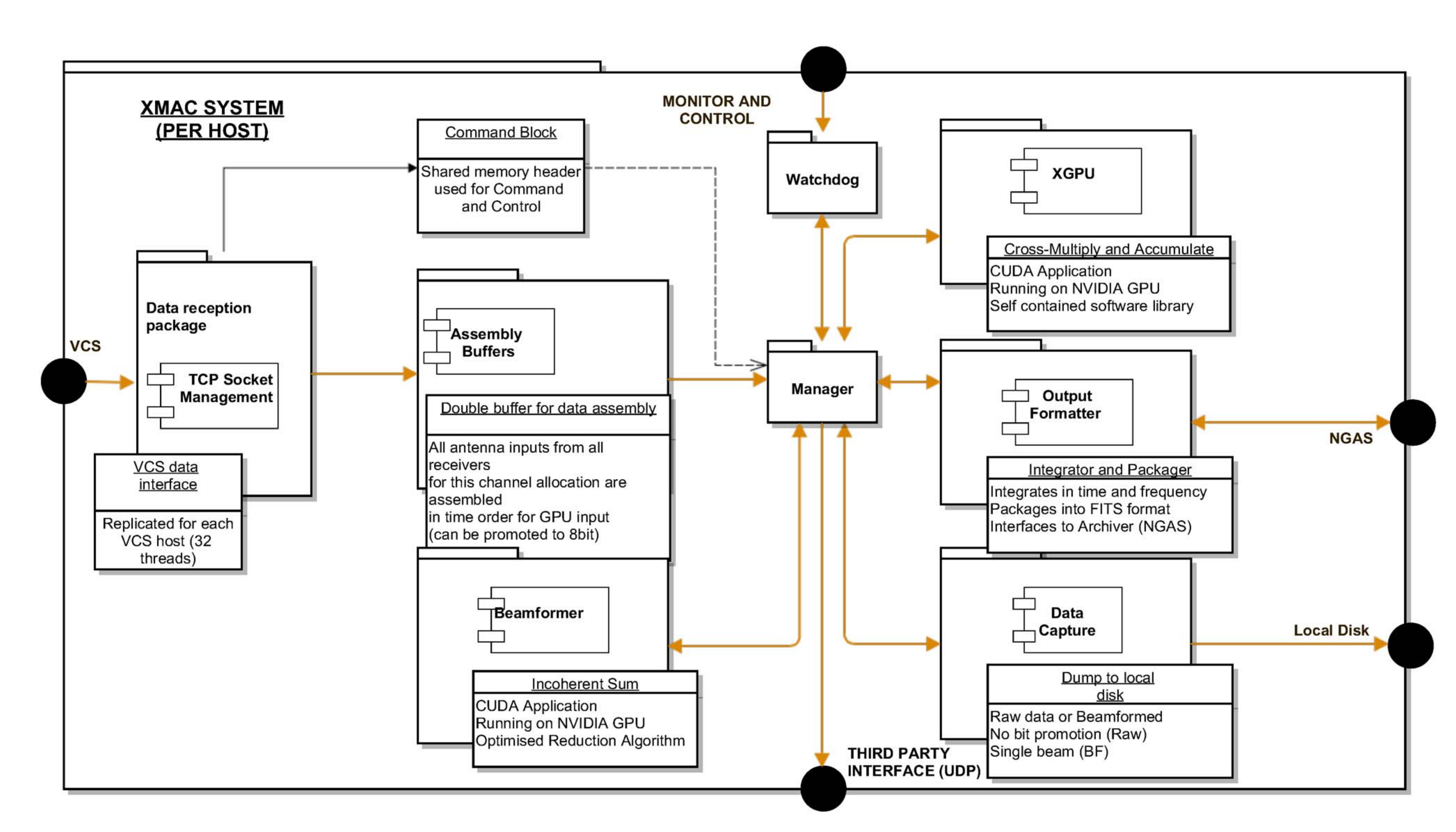}
\caption{The decomposition of the cross-multiply and accumulate operation running on each of 24, GPU enabled IBM iDataplex servers.}
\label{xmac}
\end{figure*}

\subsubsection{Input Data Reception, and Management}

A hierarchical ring buffer arrangement is employed to manage the data flow. Each TCP connection is managed by a thread that receives a 2000 packet block and places it in a shared memory buffer. Exactly where in a shared memory block each of the 2000 packets is placed is a function of the time/frequency/antenna block that it represents. A small {\em corner-turn}, or transpose, is required as the PFB mixes the order of time and frequency. The final step in this process is the promotion of the sample from 4 to 8 bits, which is required to enable a machine instruction that performs rapid promotion from 8 to 32 bits within the GPU. This promotion is not performed if the data are just being output to local disk after reordering.

Each of the 32 threads is filling a shared memory block that represents 1 second of GPU input data, and a management thread is periodically checking this block to determine whether it is full. Safeguards are in place that prevent a thread getting ahead of its colleagues or overwriting a block. The threads wait on a condition variable when they have finished their allocation, so if a data line is not getting sufficient CPU resources it will soon be the only thread {\em not} waiting and is guaranteed to complete. This mechanism helps the Linux thread scheduler to divide the resources equitably. The use of TCP allows the receiver to actually prioritise which threads on the send side of the connection are blocked to facilitate an even distribution of CPU resources.

Once the management thread judges the GPU input block to be full, it releases it, launches the GPU kernel and frees all the waiting threads to fill the next buffer while the GPU is running. 

\subsubsection{xGPU - Correlation on a GPU}

It is possible to address the resources of a GPU in a number of ways, utilising different application programming interfaces (API), such as OpenCL, CUDA, and openGL. The MWA correlator uses the xGPU library  as is described in detail in \cite{clark:2011}. The xGPU library is a CUDA application which is specific to GPUs built by the NVIDIA corporation. There are many references in the literature to the CUDA programming model and examples of its use \citep{cuda:2010}. 

The performance improvement seen when porting an application to a GPU is in general due to the large aggregate FLOPS and memory bandwidth rates, but to realise this performance requires effective use of the large number of concurrent threads of execution that can be supported by the architecture. This massively parallel architecture is permitted by the large number of processing cores on modern GPUs. The TESLA M2070 are examples of NVIDIA Fermi architecture and have 448 cores, grouped into 14 {\em streaming multiprocessors} or SM, each with 32 cores. The allocation of resources follows the following model: threads are grouped into a {\em thread block} and a thread block is assigned to an SM; there can be more thread blocks than SM, but only one block will be executing at a time on any SM; the threads within each block are then divided into groups of 32, (one for each core of the associated SM), and this subdivision called a {\em warp}; execution is serialised within a warp with the same instruction being performed by all 32 threads, but on different data elements in the {\em single instruction multiple data} paradigm. 

GPGPU application performance is often limited by memory access
bandwidth. A good predictor of algorithm performance is therefore
{\em arithmetic intensity}, or the number of floating point operations per byte
transferred. A complex multiply and add for two dual polarisation antennas
at 32 bit precision requires 32 bytes of input, 32 FLOPS (16
multiply-adds), and 64 bytes of output. The arithmetic intensity of this
is 32/96=0.33.  Compare this to the Fermi C2070 architecture which has a
peak single-precision performance of 1030 GFLOPS and memory bandwidth of
144 GB/s; the ratio of which (7.2) tells us that the performance of the
algorithm will be completely dominated by memory traffic. The high
performance kernel developed by Clark et al. (2011) increases the
arithmetic intensity in two ways. Firstly the output memory traffic can be
reduced by integrating the products for a time, I, at the register level.
Secondly, instead of considering a single baseline, groups of ($m \times n$)
baselines denoted tiles (see Figure 6), are constructed cooperatively
by a block of threads. In order to fill an $m \times n$ region of the final
correlation matrix, only two vectors of antennas need be transferred from
host memory. A thread block loads two vectors of antenna samples (of
length $n$ and $m$) and forms $nm$ baselines. These two steps then alter the
arithmetic intensity calculation to


\begin{eqnarray}
\mathrm{Arithmetic\,\,Intensity} & = & \frac{32mnI}{16(m+n)I + 64mn} \\
\end{eqnarray}

\noindent which implies that the arithmetic intensity can be made arbitrarily large by increasing the tile size until the number of available registers is exhausted. In practice a balance must be struck between the available resources and the arithmetic intensity and this balance is achieved in the xGPU implementation by {\em multi-level tiling} and we direct the reader to \cite{clark:2011} for a complete discussion.

\begin{figure}

\includegraphics[width=\columnwidth]{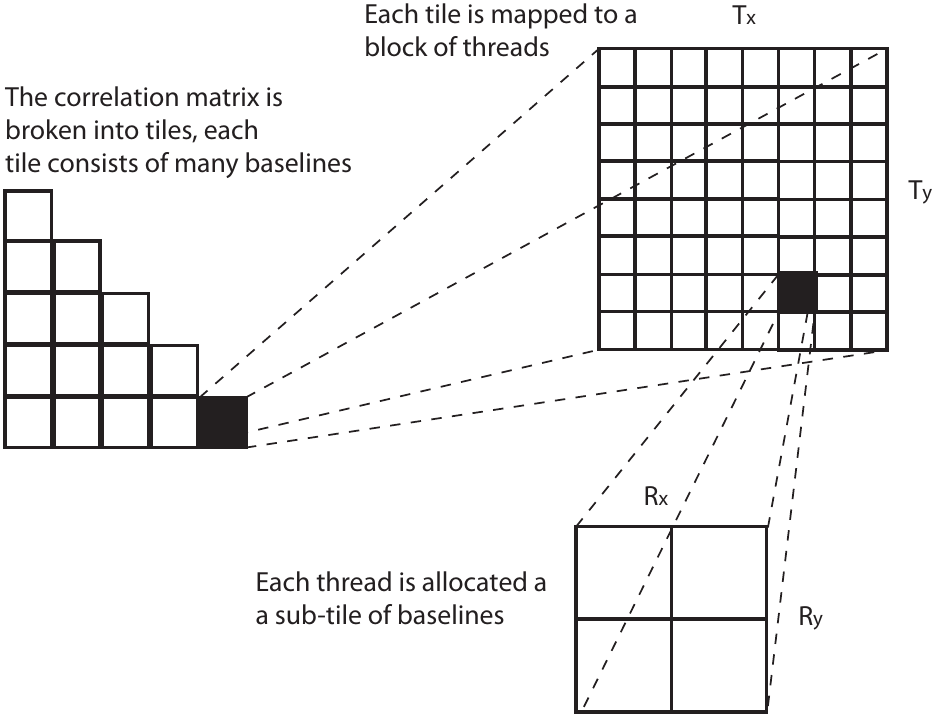}
\caption{The arithmetic intensity of the correlation operation on the GPU is increased by tiling the correlation matrix. Threads are assigned groups of baselines instead of a single baseline. }
\label{fig:mapping}
\end{figure}

\begin{figure*}
\includegraphics[width=\textwidth]{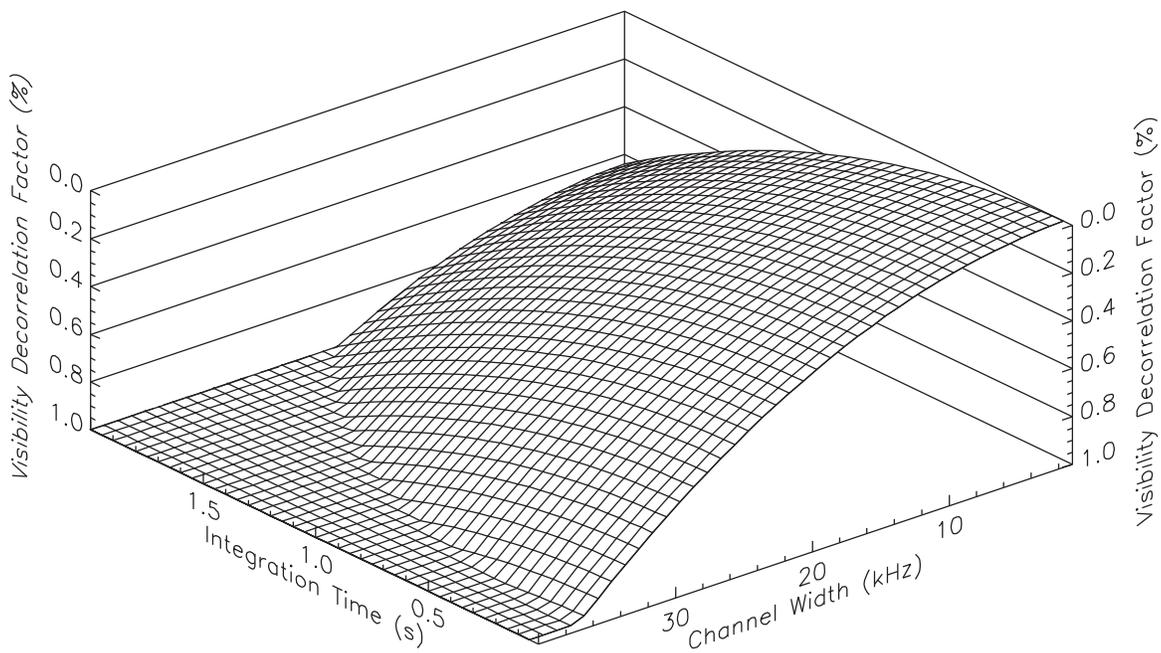}
\caption{The combined reduction in coherence due to time and bandwidth smearing 45 degrees from zenith on a 1km baseline, at an observing frequency of 200MHz as a function of channel width and integration time, demonstrating that even this far from zenith the decorrelation is generally less the 1\%. The longest MWA baselines are 3km, and these baselines show closer to 5\% decorrelation for the same observing parameters. Decorrelation factors above 1\% are not plotted for clarity.}
\label{decor}
\end{figure*}

The MWA correlator is not a delay tracking correlator. Therefore, we do not have to adjust the correlator inputs for whole, or partial sample delays. The correlator always provides correlation output products phased to zenith. The system parameters (integration time, baseline length, channel width) have all been chosen with this in mind and the long operating wavelengths result in the system showing only minimal (~1\%) decorrelation even far from zenith for typical baseline lengths, integration times and channel widths (see Figure \ref{decor}). Typical observing resolutions have been between 0.5 and 2 seconds in time and 40 kHz and 10 kHz in frequency. 

\subsubsection{Output Data Format}

The correlation products are accumulated on the GPU device and then copied asynchronously back to host memory. The length of accumulation is chosen by the operator. Each accumulation block is topped with a FITS header \citep{FITS:1981} and transferred to an archiving system where the products are concatenated into larger FITS files with multiple extensions, wherein each integration is a FITS extension (see Appendix \ref{NGAS}). The archiving system is an instantiation of the Next Generation Archiving System (NGAS) \citep{wk07b,wicenec:2013}  and was originally developed to store data from the European Southern Observatory (ESO) telescopes. The heart of the system is a database that links the output products with the meta-data describing the observation and that manages the safe transportation and archiving of data. In the MWA operating paradigm all visibilities are stored, which at the native resolution is 33024 visibilities and 3072 channels or approximately 800\,MBytes per time resolution unit (more than 6\,Gb/s for 1 second integrations). Due to the short baselines and long wavelengths used by the array it is reasonable to integrate by a factor of two in each dimension (of time and frequency, see Figure \ref{decor} for the parameter space), but even this reduced rate amounts to 17 TBytes/day at 100\% duty cycle. The large data volume precludes handling by humans and the archiving scheme is entirely automated. NGAS provides tools to access the data remotely and provides the link between the data products to the observatory SQL database. As discussed in \citet{tingay:2013}, the data are taken at the Murchison Radio Observatory in remote Western Australia and transferred to the Pawsey HPC Centre for SKA Science in Perth, where 15~PB of storage is allocated to the MWA over its five-year lifetime. Subsequently the archive database is mirrored by the NGAS system to other locations around the world: MIT in the USA; The Victoria University of Wellington, NZ; and the Raman Research Institute in India. These remote users are then able to access their data locally.

\section{Instrument Verification and Commissioning}

After initial instrument commissioning with simple test vectors and noise inputs the telescope entered a commissioning phase in September 2012. The commissioning team initially consisted of 19 scientists from 10 institutions across 3 countries and was led by Curtin University. This commissioning phase was successfully concluded in July 2013 giving way to ``Early Operations'' and  the MWA is now fully operational.  

\subsection{Verification}

The development of the MWA correlator proceeded separately to the rest of the MWA systems and the correlator GPU based elements were verified against other software correlator tools. The MWA correlator is comparatively simple, it does not track delays, performs the correlation at 32-bit floating point precision, and does not have to perform any Fourier transformations. The only verification that was required was to ensure that the cross multiply and accumulate was performed to sufficient precision and matched results generated by a simple CPU implementation.

Subsequent to ensuring that the actual cross multiply and accumulate operation was being performed correctly the subsequent testing and integration was concerned with ensuring that the signal and path was maintained. This was a relatively complex operation due to the requirement that the correlator interface with legacy hardware systems. The details of these interfaces are presented in the Appendices. In Figure \ref{blines} a correlation matrix is shown, the colours representing the length of the baseline associated with that antenna pair. The following Figure \ref{CenA} shows the response of the baselines when the telescope is pointed at the radio galaxy Centaurus A, demonstrating the response of the different baseline lengths of the interferometer to structure in the source. Images of this object obtained with the MWA can be found in McKinley et al. 2013\nocite{McKinley:2013}. The arrows on Figure \ref{blines} indicate those baselines associated with the densely-packed core \citep{tingay:2013}. 

\begin{figure}
\includegraphics[width=\columnwidth]{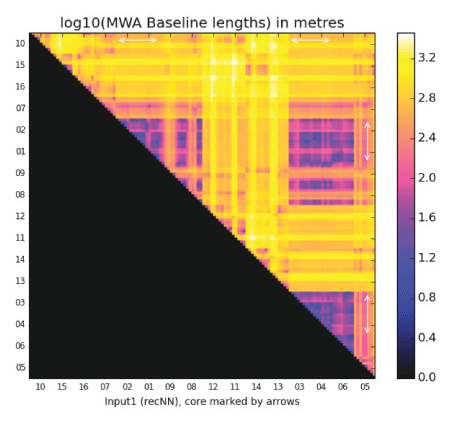}
\caption{A correlation matrix showing the baseline length distribution. The antennas are grouped into receivers, each servicing 16 antennas and the antenna layout displays a pronounced centrally dense core. The layout is detailed in \cite{tingay:2013}. The colours indicate baseline length and the core regions are also indicated by arrows within the figure.}
\label{blines}
\end{figure}

\begin{figure}
\includegraphics[width=\columnwidth]{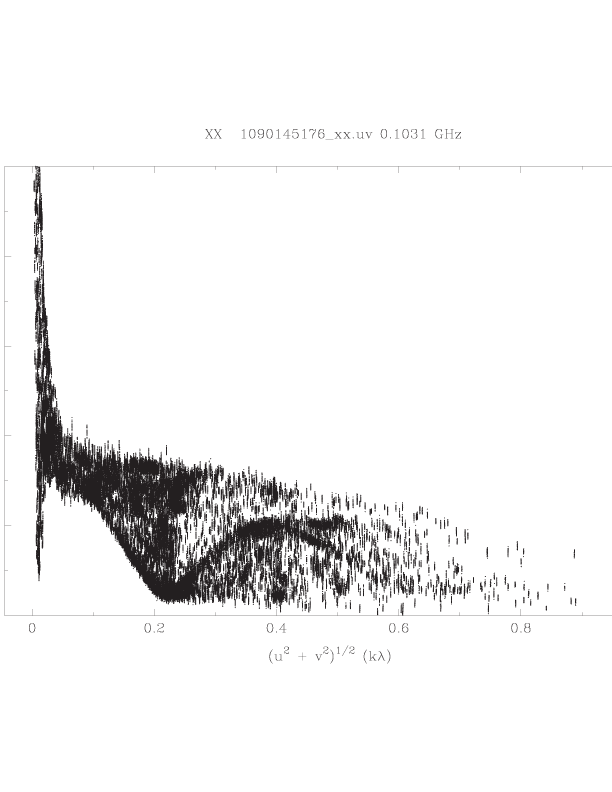}
\caption{Visibility amplitude vs baseline length for a 2 minute snapshot pointing at the giant radio galaxy Centaurus A. Centre frequency 120 MHz. Only 1 polarisation is shown for clarity. The source has structure on a wide range of scales and does not dominate the visibilities as a point source.}
\label{CenA}
\end{figure}

\subsection{Verification Experiments}

The MWA Array and correlator is also the verification platform for the Aperture Array Verification System (AAVS) within the SKA reconstruction activities pursued by the Low Frequency Aperture Array (LFAA) consortium\citep{AAVS:2011}. This group have recently performed a successful verification experiment, using both the MWA and the AAVS telescopes. Using an interferometric observation of the radio galaxy Hydra A taken with the MWA system, they have obtained a measurement of antenna sensitivity (A/T) and compared this with a full-wave electromagnetic simulation (FEKO\footnote{www.feko.info}).  The measurements and simulation show good agreement at all frequencies within the MWA observing band and are presented in detail in \cite{Colegate:2014}.

\subsection{Commissioning Science}

In order to verify the instrumental performance of the array, correlator and archive as a whole, the MWA Science Commissioning Team have performed a 4300 deg$^2$ survey, and have published a catalogue of flux densities and spectral indices from 14,121 compact sources. The survey covered approximately 21~h $<$ Right Ascension $<$ 8~h, -55$^{\circ}$ $<$ Declination $<$ -10$^{\circ}$ over three frequency bands  centred on 119, 150 and 180~MHz. This survey will be detailed in Hurley-Walker et al. (2014)\nocite{natasha:2014}. Data taken during and shortly after the commisioning phase of the instrument has also been used to demonstrate the tracking of space debris with the MWA \citep{tingay:2013b}, novel imaging and deconvolution schemes \citep{andre:2014}, and to present multi-frequency observations of the Fornax radio galaxy \citep{McKinley:2013} and the galaxy cluster A3667 \citep{hindson:2014}.


\section{The Future of the Correlator}
 
\subsection{Upgrade Path}

We do not fully utilise the GPU in the MWA correlator and we estimate that the X-Stage of the correlator could currently support a factor of 2 increase in array size (to 256 dual polarisation elements). As discussed in \cite{tingay:2013} the operational life-span of the MWA is intended to be approximately five years. The GPUs (NVIDIA M2070s)  the MWA correlator are from the Fermi family of NVIDIA GPU and are already almost obsolete. One huge advantage of off-the-shelf signal processing solutions is that we can easily benefit from improvements in technology. We could swap out the GPU in the current MWA cluster, replace them with cards from the  Kepler series (K20X) with no code alteration, and would benefit from a factor of 2.5 increase in performance (and a threefold improvement in power efficiency) as not only are the number of FLOPS provided by the GPUs increasing, but also their efficiency (in FLOPS/Watt) is improving rapidly. This would permit the MWA to be scaled to 512 elements. However, not all of our problem is with the correlation step and increasing the array to 512 elements would probably require an upgrade to the networking capability of the correlator. 

\subsection{SKA Activities}

The MRO is the proposed site of the low-frequency portion of the Square Kilometre Array. Curtin University is a recipient of grants from the Australian government to support the design and pre-construction effort associated with the construction and verification of the Low-Frequency Aperture Array (LFAA) and software correlation systems for the SKA project. The MWA correlator will therefore be used as a verification platform within the prototype for the LFAA, known as the Aperture Array Verification System (AAVS). One of the benefits of a software based signal processing system is flexibility. We have been able to easily  incorporate AAVS antennas into the MWA processing chain to facilitate these verification activities, which have been of benefit to both LFAA and the MWA \citep{Colegate:2014}. Furthermore we will be developing and deploying SKA software correlator technologies throughout the pre-construction phase of the SKA using the MWA correlator for both testing and verification.






\section{Summary}

This paper outlines the structure, interfaces, operations and data formats of the MWA Hybrid FPGA/GPU correlator. This system combines off-the-shelf computer hardware with bespoke digital electronics to provide a flexible and extensible correlator solution for a next generation radio telescope. We have outlined the various stages in the correlator signal path and detailed the form of all internal and external interfaces. 

\section{Acknowledgements}

We acknowledge the Wajarri Yamatji people as the traditional owners of the Observatory site.  Initial correlator development was enabled by equipment supplied by an IBM Shared University Research Grant (VUW \& Curtin University), and by an Internal Research and Development Grant from the Smithsonian Astrophysical Observatory.

Support for the MWA comes from the U.S. National Science Foundation (grants AST CAREER-0847753, AST-0457585, AST-0908884 and PHY-0835713), the Australian Research Council (LIEF grants LE0775621 and LE0882938), the U.S. Air Force Office of Scientific Research (grant FA9550-0510247), the Centre for All-sky Astrophysics (an Australian Research Council Centre of Excellence funded by grant CE110001020), New Zealand Ministry of Economic Development (grant MED-E1799), an IBM
Shared University Research Grant (via VUW \& Curtin), the Smithsonian Astrophysical Observatory, the MIT School of Science, the Raman Research Institute, the Australian National University, the Victoria University of Wellington, the Australian Federal government via the National Collaborative Research Infrastructure Strategy, Education Investment Fund and the Australia India Strategic Research Fund and Astronomy Australia Limited, under contract to Curtin University, the iVEC Petabyte Data Store, the Initiative in Innovative Computing and NVIDIA sponsored CUDA Center for Excellence at Harvard, and the International Centre for Radio Astronomy Research, a Joint Venture of Curtin University and The University of Western Australia, funded by the Western Australian State government.

\appendix

\section{MWA F-Stage (PFB) to Voltage Capture System (VCS)}
\label{VCS}
The contents of this Appendix are extended from the original MWA ICD for the polyphase filterbank written by R. Cappallo. We have incorporated the changes made to the physical mappings, electrical connections  and header contents to enable capture of the PFB output by general purpose computers.
\subsection{Physical}
\begin{itemize}
\item{PFB presents data on two CX-4 connectors. ALL pins are wired with transmit drivers, there are no receivers on any pins}
\end{itemize}
\subsection{Interface Details}
\begin{itemize}
\item{Two custom made CX4 to SFP breakout cables. This cable breaks a CX4 connector into 4 1X transmit and receive pairs on individual SFP terminated cables.}
\item{Data flows only in a single direction (simplex) from PFB to CB. Each (normally Tx/Rx) pair of differential pairs is wired as two one-way data paths (Tx /Tx). We have distributed these one-way data paths to occupy every other CX-4 pin. Therefore all cables are wired correctly as Tx/Rx but the Rx channels are not utilised.}
\item{The serial protocol to be used a custom protocol as described below. The interface to
the Xilinx Rocket I/O MGTÕs is GT11\_CUSTOM, which is a lean protocol. The word width
will be 16 bits.}
\item{The mean data rate per 1X cable will be 2.01216 Gb/s, carried on a high speed serial
channel of burst rate of 2.08 Gb/s formatted, or 2.6 Gb/s including the 10B/8B encoding
overhead.}
\end{itemize}



\subsection{Data Format}

\begin{itemize}
\item{Each data stream carries all of the antenna information for 384 fine frequency channels.
In the entire 32T correlation system there are 8 such data streams going from the PFB to
the CB, each carrying the channels making up a 3.84 MHz segment of the processed
spectrum.}
\item{ Each sample in the stream is an 8-bit (4R+4I) complex number, 2Õs complement with 8
denoting invalid data, representing the voltage in a 10 kHz fine frequency channel.}
 
 \item{Data are packetised and put in a strictly defined sequence, in f1t[f2a] (frequency-time-frequency-antenna) order. The most rapidly varying index, a, is the antenna number (0,
16, 32, 48, 1, 17, 33, 49, 2,18, É, 15, 31, 47, 63), followed by f2, the fine frequency
channel (n..n+3), then time (0..499), and most slowly the fine frequency channel group
index (0, 4, 8, 12, 128, 132, 136, 140, ,É, 2944, 2948, 2952, 2956 for fibre 0; increment
by 16n for fibre n). The brackets indicate the portion of the stream that is contained
within a single packet. The antenna ordering is non intuitive as for historical reasons each PFB combines its inputs into an order originally considered conducive to its interface to a full mesh backplane and a hardware FPGA based correlator.}
 
\item{A single packet consists of all antenna data for a group of four fine channels for a single
time point. Thus there are 500 sequential time packets for each of the 96 frequency
channel groups. Over the interval of 50 ms there are 48,000 (=96channel groups x 500 time samples) packets
sent per data stream, so there are 960,000 packets/second.}
\item{ Using a packet size of 264 bytes allows a 16-bit packet header (0x0800), two 16-bit words
carrying the second of time tick and packet counter, 256 bytes of antenna data, finally followed by a 16-bit
checksum (see Table 1).}
\item{The header word is a 16-bit field with value 0x0800, used to denote the first word in a
packet. Note that it cannot be confused with the data words, since the value 8 is not a
legal voltage sample code.}
\item{The second tick in word two is in bit 0, and has the value 0 at all times, except for the
first packet of a new second, when it is 1.}
\item{The leading bits of word two, and all of word three contains the following counter values that can be decoded to determine the precise input and channel that a given packet belongs to.
\begin{itemize}
    \item {\em sec\_tick} (1 bit): Is this the first packet for this data lane, for this one
       second of data? 0=No, 1=Yes.  Only found on the very first packet of that
       second.  If set, then the mgt\_bank, mgt\_channel, mgt\_group and mgt\_frame
       will always be 0. The preceding 3 bits are unused.

   \item {\em pfb\_id} (2 bits): Which physical PFB board generated this stream [0-3].
       This defines the set of receivers and tiles the data refers to.
       A lane's pfb\_id will remain constant unless physically shifted
       to a different PFB via a cable swap.

   \item {\em mgt\_id} (3 bits): Which 1/8 of coarse channels this Rocket--IO lane contains.
       Should be masked with 0x7, not 0xf.  Contains [0-7] inclusive.
       A lane's mgt\_id will remain constant unless physically shifted
       to a different port on the PFB via a cable swap.
\end{itemize}
   Within an individual data lane, the data packets cycle in the following order, listed from
       slowest to fastest.
\begin{itemize}
 \item  {\em mgt\_bank} (5 bits): Which one of the twenty 50ms time banks in the current
        second is this one? [0-19] [0-0x13].

  \item {\em mgt\_channel} (5 bits): Which coarse channel [0-23] [0-0x17] does the packet
       relate to?

  \item  {\em mgt\_group} (2 bits): Which 40KHz wide packet of the contiguous 160KHz does
        this packet contain the 4 x 10KHz samples for? [0-3]

   \item {\em mgt\_frame} (5 bits): Which time stamp within a 50ms block this packet is.
       [0-499] or [0-0x1F3].  Cycles fastest.  Loops back to 0 after 0x1F3.
       There are 20 complete cycles in a second.
       \end{itemize}
       }

\item{The last word of the packet is a 16-bit checksum, formed by taking the bitwise XOR of
all 128 (16-bit) words of antenna samples in the packet.}
\end{itemize}



\section{VCS to XMAC Server}
\label{XMAC_Ap}
\subsection{Physical}
\begin{itemize}
\item{Data presented on either fibre optic or direct attach copper cables terminated with SFP pluggable transceivers. Note that all optical transceivers used in the CISCO UCS servers must be supplied by CISCO, the firmware within the 10Gb ethernet cards requires it. }
\item{A single 10GbE interface is sufficient to handle the data rate}
\end{itemize}
\subsection{Interface Details}

\begin{itemize}
\item{Ethernet IEEE 802.3 frames, with TCP/IPv4}
\item{The packet format is precisely as described in the PFB to VCS interface}
\item{The communication is from the VCS to the XMAC is uni-directional and mediated by a switch.}
\item{Each XMAC requires a data packet from all VCS machines. This requires 32 open connections mediated by TCP}
\item{Data from all sources is stripped of header information and assembled into a common 1 second buffer which requires 20 blocks of 2000 packets from each of the 32 connections to be assembled.}
\item{As the packets for four PFB are being combined the antenna ordering is also being concatenated. The order within each packet is maintained with the 64 inputs concatenated into a final block of 256 inputs.}
\end{itemize}

\section{XMAC to GPU Interface}
\label{GPU_Ap}
\subsection{Physical}
The interface is internal to the CPU host, being a section of memory shared between the GPU device driver and the host.

\subsection{Interface details}

\subsubsection{GPU Input}
The data ordering in the assembly buffer for input buffers, running from slowest to fastest index is time(t), channel(f), station(s), polarisation(p), complexity(i). The data is an 8 bit integer, promoted from 4-bit twos-complement sample via a lookup table. The correlator is a dual polarisation correlator as implemented, but the input station/polarisation ordering is not in a convenient order as we have concatenated 4 PFB outputs during the assembly of this buffer. Two options presented themselves. One that we re-order the input stations and polarisation to undo the partial corner turn. The disadvantage begin that each input packet would need to be broken open and the stations reordered - which would have been extremely compute intensive. Or to correlate {\em without} changing the order and remapping the output products to the desired order. This is preferable as although there are now N$^2$ products instead of N stations, the products are integrated in time and can be reordered in post-processing.

\subsubsection{GPU Output}
The data are formated as 32-bit complex floating point numbers. An individual visibility set has the order (from slowest index to fastest index): channel(f), station(s1), station(s2), polarisation(p1), polarisation(p2), complexity(i). The full cross-correaltion matrix is Hermitian so only the lower, or upper, triangular elements need to be calculated, along with the on-diagonal elements which are the auto-correlation products. Therefore only a packed triangular matrix is actually transferred from the GPU to the host. The GPU processing kernel is agnostic to the ordering of the antenna inputs and the PFB output order is maintained. Therefore the correlation products cannot be simply processed without a remapping.

The remapping is performed by a utility that has been supplied by the builders to the data analysis teams that simply reorders the data products into one  that would be expected if no PFB reordering had taken place. It is important to keep track of which products have been generated but the XMAC as conjugates of the desired products, and which need conjugating. It should further be noted that this correction is performed to the order as presented {\em to} the PFB and may not the order as expected by the ordering of the receiver inputs. Care should be taken in mapping antennas on the ground as the order of the receiver processing within a PFB is not only function of the physical wiring but the firmware mapping. We have carefully determined this mapping and the antenna mapping tables are held within the same database as the observing metadata.

For historical reasons the files are transferred into an intermediate file format that was originally used during instrument commissioning before subsequent transformation to UVFITS files (or measurement sets) by the research teams. 

\section{XMAC Server to NGAS (Next Generation Archiving System)}  
\label{NGAS}
\subsection{Physical}
The interface is internal to the CPU host, being a section of memory shared between the XMAC application and an NGAS application running on the same host. 

\subsection{Interface details}
The data are handed over as a packed triangular matrix with the same ordering as was produced by the GPU. It has a FITS header added and is padded to the correct size as expected by a FITS extension. The FITS header includes a time-tag.

Once the buffer is presented to the NGAS system it is buffered locally and presented to a central archive server, subsequently it is added to a database and mirrored to multiple sites across the world.

\bibliographystyle{apj}
\bibliography{journals}


\end{document}